**Scalable, 'Dip-and-dry' Fabrication of a Wide-Angle Plasmonic Selective Absorber for High-efficiency Solar-thermal Energy Conversion**

*Jyotirmoy Mandal, Derek Wang, Adam C. Overvig, Norman N. Shi, Daniel Paley, Amirali Zangiabadi, Qian Cheng, Katayun Barmak, Nanfang Yu\*, Yuan Yang\**

J. Mandal, A. Overvig, N. Shi, Dr. A. Zangiabadi, Dr. Q. Cheng, Prof. K. Barmak, Prof. N. Yu*, Prof. Y. Yang*
Department of Applied Physics and Applied Mathematics, Columbia University,
Mudd 200, MC 4701, 500 W 120 Street, New York, NY 10027, U.S.A.
Email: yy2664@columbia.edu, ny2214@columbia.edu
Dr. D. Paley
Department of Chemistry, Columbia University,
3000 Broadway New York, NY 10027
D. Wang
Department of Materials Science and Engineering, Durand Building, Stanford University
496 Lomita Mall, Stanford, CA 94305-4034, U.S.A.

*Abstract for the reviewer:* A galvanic displacement reaction-based, room-temperature 'dip-and-dry' technique is demonstrated for fabricating selectively solar-absorbing plasmonic nanostructure-coated foils (PNFs). The technique, which allows for facile tuning of the PNFs' spectral reflectance to suit different radiative and thermal environments, yields PNFs which exhibit excellent, wide-angle solar absorptance (0.96 at 15°, to 0.97 at 35°, to 0.79 at 80°) and low hemispherical thermal emittance (0.10) without the aid of antireflection coatings. The thermal emittance is on par with those of notable selective solar absorbers (SSAs) in the literature, while the wide-angle solar absorptance surpasses those of previously reported SSAs with comparable optical selectivities. In addition, the PNFs show promising mechanical and thermal stabilities at temperatures of up to 200°C. Along with the performance of the PNFs, the simplicity, inexpensiveness and environment-friendliness of the 'dip-and-dry' technique makes it an appealing alternative to current methods for fabricating selective solar absorbers.

In recent decades, the increasing global energy expenditure and concerns about its environmental impact has seen the world significantly shift towards using renewable energy.[1] Of the different sources of renewable energy, solar radiation is the most abundant and accessible – the annual solar energy incident on the Earth's surface is ~1 x $10^4$ times the current global energy use, and even highly conservative estimates of the harvestable solar energy are predicted to meet global energy consumptions until the end of this century.[2] Solar energy can be harvested by various means, such as conversion to electricity by photovoltaic devices, to chemical energy in fuels (e.g., $H_2$), and to heat by photo-thermal converters.[3-7] The last is particularly promising, as it typically utilizes a larger bandwidth within the solar spectrum, and attains higher energy-conversion efficiencies.[8-10] Consequently, in recent years, solar-thermal converters have seen increasing uses in a variety of applications. For instance, low-temperature ($\lesssim$ 100°C) solar-thermal converters are used for passive heating, as well as distilling or desalinating water for industrial and domestic use,[11-14] while mid- (100-400°C) and high- (> 400°C) temperature variants have additional uses in concentrated solar power systems and solar-thermoelectric generators.[10, 15] Regardless of its type, an ideal solar-thermal converter maximizes its radiative heat gain from its surroundings, i.e., it has a surface with high absorptance in the solar wavelengths (0.3 to ~2.5 μm) and low emittance in the infrared thermal radiation wavelengths (~ 2.5 to 40 μm). Selective solar absorbers (SSAs), which exhibit such optical properties, have been developed for decades. However, producing low-cost selective absorbers with both high solar-thermal conversion efficiency and high durability remains challenging.

In this paper, the authors demonstrate a simple, room-temperature, 'dip-and-dry' technique to fabricate a class of plasmonic metal nanoparticle-based SSAs with high performance and high stability at temperatures of up to 200°C. The process is based on a galvanic displacement reaction, and yields plasmonic nanoparticle-coated metal foils (PNFs) with an excellent wide-angle solar absorptance that peaks at 0.97 and remains as high as 0.79 at a grazing incidence angle of 80°. The hemispherical thermal emittance is measured to be 0.10 at 100°C. The solar absorptance and thermal emittance are attained without the use of antireflection coatings, and are easily tuneable to suit different operating conditions. Furthermore, the technique does not require electrochemical or vacuum deposition techniques, and is fully compatible with roll-to-roll processes – making it a simple, inexpensive and environment-friendly alternative to current SSA manufacturing methods.

The performance of an SSA is determined by multiple factors – in particular, the intensity and angle of the incident sunlight, the device architecture and operating temperature. For maximum solar-thermal conversion efficiency, an SSA must maximize its directional solar absorptance $\bar{\alpha}(\theta)$ and minimize its hemispherical thermal emittance $\bar{\epsilon}$, which are defined as:

$$\bar{\alpha}(\theta) = \frac{\int_0^\infty I_{solar}(\lambda) . \alpha_{solar}(\theta,\lambda) d\lambda}{\int_0^\infty I_{solar}(\lambda) d\lambda} \quad (1)$$

$$\text{and } \bar{\epsilon} = \frac{\int_0^\infty I_{BB}(T,\lambda) . \epsilon_{thermal}(T,\lambda) d\lambda}{\int_0^\infty I_{BB}(T,\lambda) d\lambda} \quad (2)$$

where $\theta$ is the angle from the normal, $\lambda$ is the wavelength, $\alpha_{solar}(\theta,\lambda)$ and $\epsilon_{thermal}(T,\lambda)$ are the spectral directional solar absorptance and hemispherical thermal emittance of the SSA, $I_{solar}(\lambda)$ is the AM 1.5 Global solar intensity spectrum, and $I_{BB}(T, \lambda)$ is the spectral intensity emitted by a blackbody at temperature $T$. For opaque objects, both $\alpha_{solar}(\theta,\lambda)$ and $\epsilon_{thermal}(T,\lambda)$ can be expressed as 1 – $R(\lambda)$, where $R(\lambda)$ is the spectral reflectance. As shown in **Figure 1**, $R(\lambda)$ is ideally a step function spanning the ultraviolet to far-infrared wavelengths, with a transition wavelength that varies with the operating temperature of the SSA – for 100-200°C, it is ~2.5 μm. In this paper, an operating temperature of 100°C is assumed, and unless stated otherwise, $\bar{\alpha}$, $\bar{\epsilon}$ and associated values are quoted for near-normal incidence ($\theta = 20°$).

Research efforts in recent decades have explored different designs for achieving selective solar absorption. Variants of such designs include intrinsic selective absorbers, semiconductor-metal tandems, metal-dielectric multilayer broadband absorbers, textured metals, ceramic-metal composites (cermets), and photonic crystals.[16-20] While intrinsic absorbers rely on material properties such as interband transitions (as in W) and lowered plasma frequencies (as in $MoO_3$ doped Mo) to achieve selective solar absorption,[21] the other variants all rely on structure to enhance the intrinsically high $\bar{\alpha}$ and low $\bar{\epsilon}$ of their material constituents. For instance, cermet SSAs have metal nanoparticles embedded in a dielectric – the broad plasmon resonances of the metal nanoparticles in the solar wavelengths lead to a high $\bar{\alpha}$, while the low emittances of the dielectric and an underlying metal result in a low $\bar{\epsilon}$.[19, 21-22] Photonic crystals are more sophisticated, and use periodic arrangements of dielectric or metallic structures to suppress reflection and trap sunlight to enhance selective solar absorption.[20]

The aforementioned designs are all known or predicted to have values of $\bar{\alpha}$ ($\gtrsim 0.8$) and $\bar{\epsilon}$ ($\lesssim 0.2$), as required for practical use.[10] However, with regard to cost, ease of fabrication or environmental footprint, they may be less than desirable. Fabricating multilayer structures, for instance, involves expensive vacuum-based vapor-deposition processes,[16] while making cermets may require hazardous chemicals.[18] A scalable and environment-friendly process to develop efficient, durable and low-cost SSAs remains sought-after. Towards this end, a room-temperature, galvanic-displacement reaction based 'dip-and-dry' technique can provide a facile, inexpensive and environment-friendly alternative for fabricating high-performance SSAs.

Plasmonic metal nanoparticles are attractive for selective solar absorption because they can effectively scatter, trap and absorb radiation in the solar wavelengths.[23] The plasmon resonances induced by sunlight undergo non-radiative damping, generating heat in the process. Typically, plasmonic nanoparticles are synthesized as dispersions in liquids by chemical methods, which makes deposition on substrates complicated. However, under suitable conditions, nanoparticles of certain metals M1 can be easily formed from a solution of M1$^{x+}$ ions on another metal M2 by galvanic displacement reaction. The nanoparticles of M1, which are plasmonic, form a sub-micron layer with a high solar absorptance. However, at longer, infrared wavelengths corresponding to thermal radiation (4-20 μm), the nanoparticles act as a lossy effective medium with deep-subwavelength thickness on the highly reflective M2, leading to a low thermal emittance. Here, the authors present copper (Cu) and zinc (Zn) as one such M1-M2 pair (further examples are described in the Supporting Information (SI), Section 6). Figure 1 summarizes the fabrication process and the characteristics of a Cu-Zn PNF. By merely dipping Zn foil in aqueous $CuSO_4$ for ~30-60s, aqueous $Cu^{2+}$ ions can be reduced to metallic Cu nanoparticles by Zn on its surface (Figure 1a-b). The Cu deposition appears as a black layer on the metallic Zn (Figure 1c), indicating a strong solar absorptance (Figure 1d). Visible and infrared reflectance measurements reveal that the PNF has an excellent optical selectivity ($\bar{\alpha}$ = 0.96 and $\bar{\epsilon}$ = 0.08, Figure 1e). The performance is all the more remarkable given the PNF's simple fabrication, and on par with commercial SSAs and notable results in literature.[10, 18-19, 23-24]

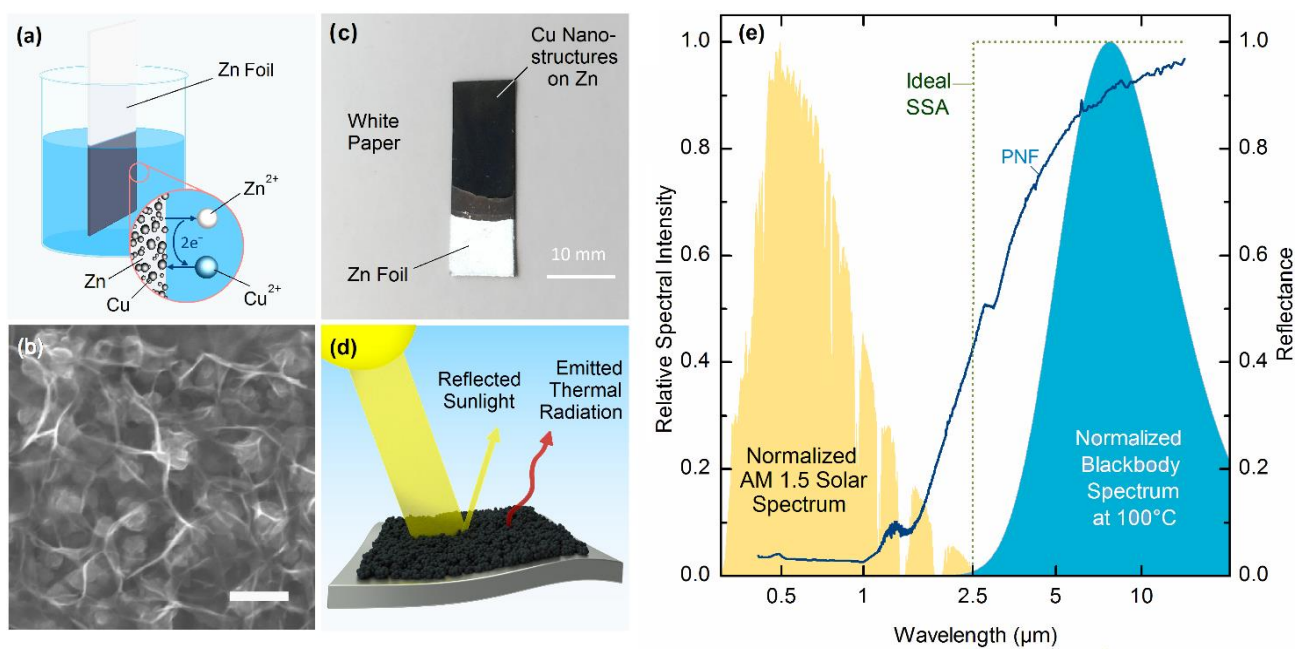

**Figure 1**. a) Schematic of the deposition process, showing the formation of Cu nanoparticles on Zn by galvanic displacement reaction. b) Scanning electron microscope (SEM) image of the Cu nanoparticle layer on the PNF. The scale bar represents a length of 200 nm. c) Photograph of the PNF against a white background, with light reflecting specularly off the sample into the camera. d) Schematic depicting the high solar absorptance and low thermal emittance of the PNF. Thickness of the arrows indicate intensity. e) Spectral reflectance of a PNF ($\bar{\alpha}$ = 0.96, $\bar{\epsilon}$ = 0.08) and of the ideal SSA at 100°C. Normalized spectral intensities of the AM 1.5 solar spectrum and a blackbody at 100°C are also shown.

It is well-known that the Cu-Zn galvanic-displacement reaction yields metallic copper.[25] For the nanoparticle layer of the PNFs, this is confirmed by multiple characterizations shown in Figure 2. For instance, nanobeam electron-diffraction from nanoparticles extracted from the PNF shows peaks that correspond to polycrystalline, FCC copper (Figure 2b) – no other phase is observed. Elemental maps (Figure 2c-e) show a trace of Zn around the Cu nanoparticle, which is confirmed by Energy Dispersive X-ray spectroscopy as small compared to Cu (Figure 2f). This is further corroborated by X-ray diffraction (XRD) measurements of the nanoparticles, which show a pure copper phase (Figure 2g), and by Auger electron spectroscopy (AES) of the PNF, which show pure, elemental Cu 20 nm below the PNF's surface (SI, Figure S11b). The corroborating characterizations firmly establish the nanoparticle layer as metallic, and confirms the PNF's plasmonic behavior.

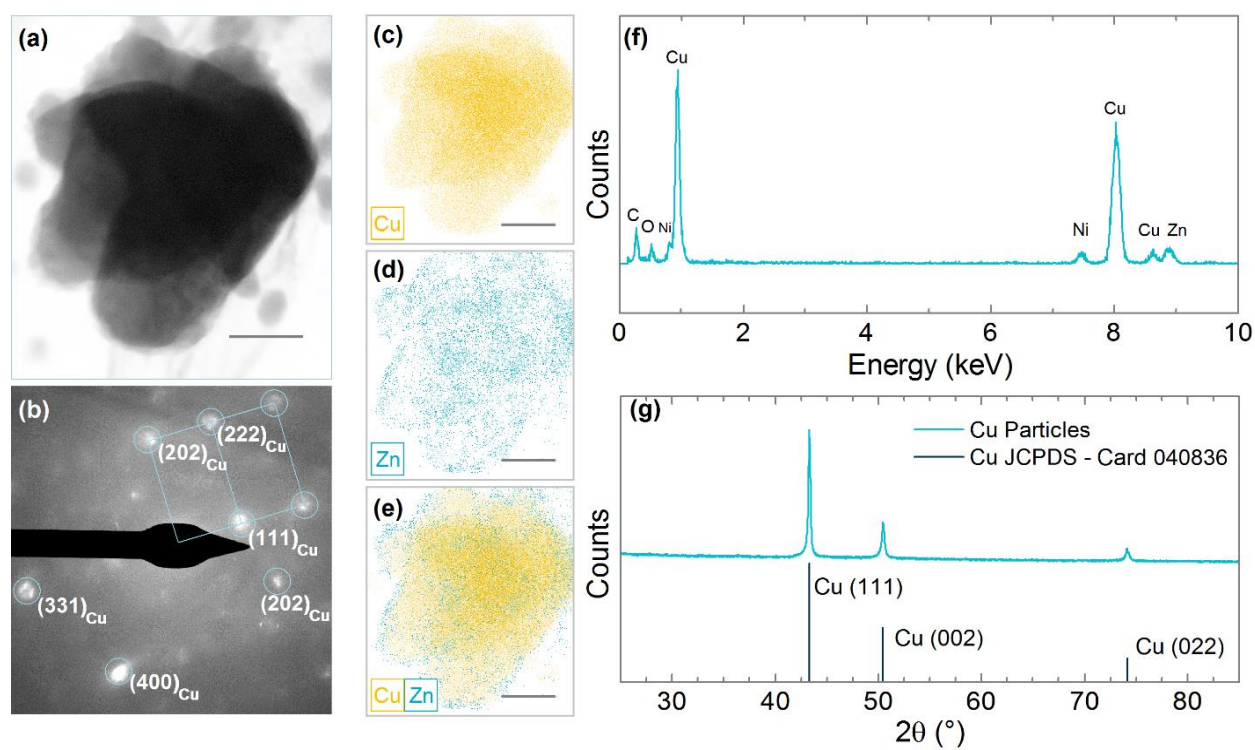

**Figure 2.** Characterizations of particles extracted from the PNF's surface (as described under 'Material Characterizations' in the Experimental Section). a) Transmission electron microscope (TEM) image of a Cu nanoparticle. Scale bar represents a length of 50 nm. b) Nanobeam diffraction pattern of the nanoparticle in (a), showing strong peaks (circled) corresponding to Cu. Overlapping patterns of the peaks suggest that the Cu is polycrystalline. c) Cu, d) Zn, and e) combined elemental maps of the nanoparticle in (a) obtained using the TEM's scanning (STEM) mode. The maps show a thin layer of Zn around the nanoparticle. Scale bar represents a length of 50 nm. f) EDS spectra of the nanoparticle in (a). The peaks corresponding to Nickel (Ni) originate from the nickel mesh supporting the samples. g) X-ray diffraction patterns of Cu particles from the PNF, and a JCPDS Cu reference. Zn is not observed because the particles are extracted from the PNF to eliminate the overwhelming signature from the Zn substrate. As evident, the characterizations all confirm the presence of metallic Cu nanoparticles on the PNF.

For plasmonic nanoparticles, light absorption depends on factors such as size, material environment, and total volume of the absorbing metal.[26] Therefore, to understand the mechanisms behind the spectral selectivity of the PNF and predict its optimal morphology for selective solar absorption, different morphologies of the PNF and their spectral reflectances were simulated using the finite-difference time-domain technique (Lumerical FDTD Solutions). As shown in **Figure 3a**, the Zn substrate was represented by an optically thick slab, on which Cu nanoparticles were randomly arranged. Each Cu nanoparticle was represented as a cluster (right inset, Figure 3a) of small, 20 nm spheres to mimic the surface features of the nanoparticles (left inset, Figure 3a). The diameters (*d*) of the nanoparticles and the thickness (*h*) of the nanoparticle layer were varied, and spectral reflectances at normal incidence for the 400 nm-14 μm wavelengths were recorded. As shown in Figure 3b-d, with increasing layer thickness, solar absorptance $\bar{\alpha}$ and thermal emittance $\bar{\epsilon}$ both rise. This can be attributed to several effects. In the solar wavelengths, the high absorptance arises from the plasmon resonances of the Cu nanoparticles. The spectral absorbance is broader than that of bulk copper, likely due to resonance-broadening arising from electron scattering at the boundaries of and defects within the nanoparticles.[7] As *h* increases, the volume for both light-matter interactions and near-field coupling between neighboring nanoparticles increase[24] – the first strengthens the absorption, and the second broadens it. On the other hand, at thermal radiation wavelengths $\gg d$, the nanoparticle layer behaves as a lossy effective medium with low reflectivity, which light traverses before and after it is reflected by the Zn substrate. Therefore, increasing thickness leads to a higher attenuation of the light and thus a lower reflectance (SI, Figure S2). A broader absorption (i.e. lowered reflectance) extending into the infrared wavelengths due to resonance peaks is also seen with increasing *d*, (Figure 3b) although simulations indicate that *h* has a stronger effect (SI, Figure S1). The broadening with *d* is similar to earlier reports in literature, and is likely due to higher-order multipolar resonances that manifest with increasing particle size.[26-27]

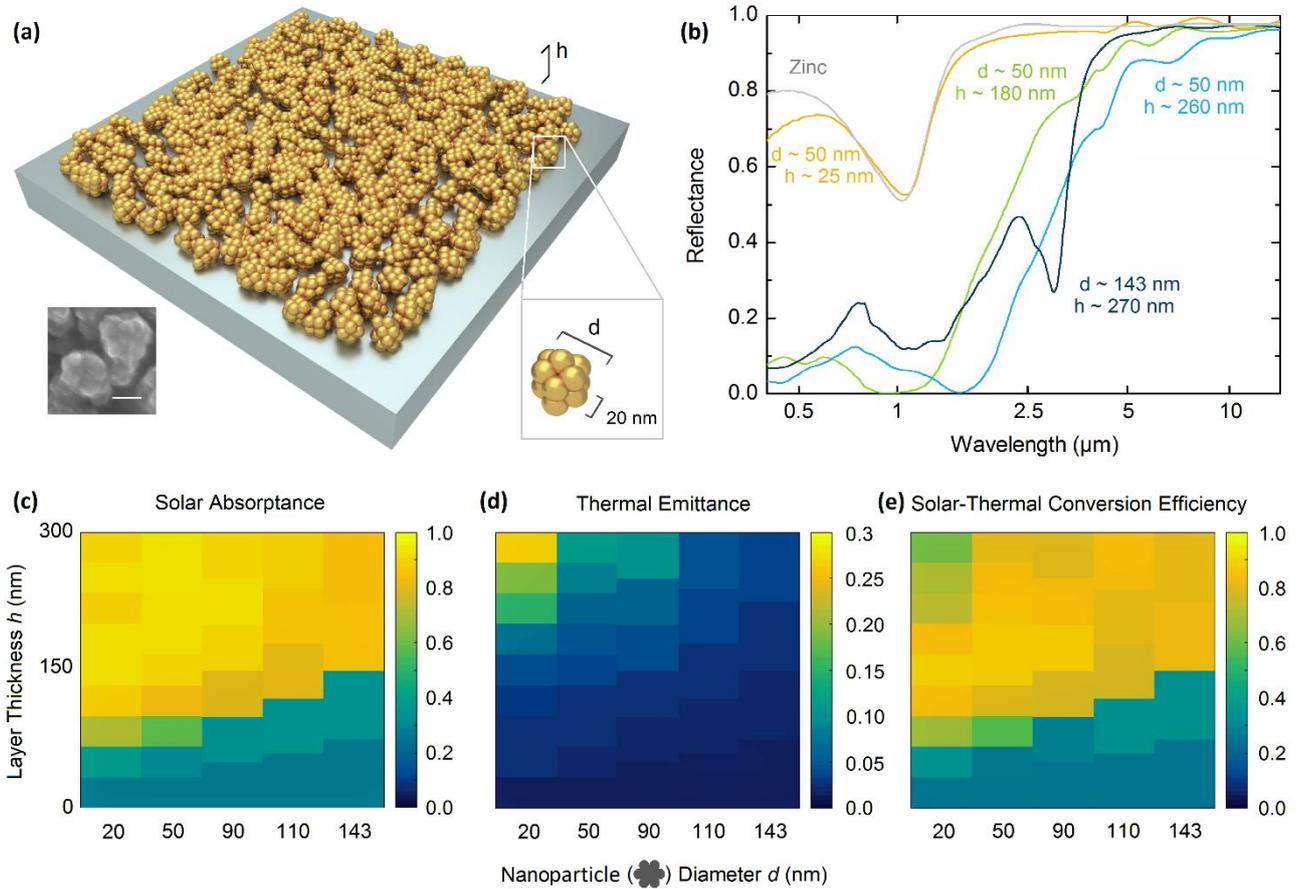

**Figure 3.** a) Schematic of the simulated PNF. The SEM image in the left inset shows Cu nanoparticles comprising of smaller features. The scale bar represents a length of 50 nm. The right inset shows a single simulated spherical nanoparticle as a cluster of 20 nm spheres. b) Simulated spectra showing the effect of increasing nanoparticle diameter *d* and layer thickness *h*. c), d) and e) respectively show $\bar{\alpha}$, $\bar{\epsilon}$ and $\eta_{sol\text{-}th}$ at normal incidence for different values of *d* and *h*. Cases with $h < d$ correspond to a single layer of 'half-nanoparticles' on the Zn.

The plots in Figure 3c-d of $\bar{\alpha}$ and $\bar{\epsilon}$ at normal incidence show general trends, which yield estimates of what nanoparticle diameters and layer thicknesses optimize the PNF's solar absorptance and thermal emittance. The absorptance and emittance can be used to derive a normal-incidence solar-thermal conversion efficiency $\eta_{sol-th}$, defined as:

$$\eta_{sol-th} = \frac{I_{solar,0°} \times \bar{\alpha} - I_{BB,100°C} \times \bar{\epsilon}}{I_{solar,0°}} \quad (3)$$

where $I_{solar,0°} = 1017$ W m$^{-2}$ is the solar intensity obtained from a normal-incidence correction of the AM 1.5 Solar Spectrum (SI, Section 3) and $I_{BB,\,100°C}$ is the emitted power from a 1 m$^2$ blackbody at 100°C. The plot of $\eta_{sol\text{-}th}$ in Figure 3e shows that diameters $d < 100$ nm and layer thicknesses $h \sim 150\text{-}300$ nm are optimal for selective solar absorption, and yield $\eta_{sol\text{-}th} \sim 0.85$. Accounting for hemispherical thermal emittance, which is typically higher than near-normal emittance,[10] the real efficiency is predicted to peak between 0.8 and 0.9.

The simulation results provide guidance on the design of a PNF for selective solar absorption. It was conjectured that the desired ranges for nanoparticle size $d$ and layer thickness $h$ can be attained by varying the galvanic-displacement reaction parameters, such as Cu$^{2+}$ concentration, immersion time, and temperature. **Figure 4** summarizes the experimental results. As evident from Figure 4a-c, the spectra closely resemble the simulation results in Figure 3b, especially considering that the fabricated PNFs exhibit spectral features corresponding to distributions of nanoparticle diameters (Figure 4g-h). Figure 4a-c show that higher reactant concentrations, longer immersion times and higher temperatures all result in lower reflectances across larger bandwidths. For immersion time, this manifests as a near constant $\bar{\alpha}$ (~0.83) and a small rise in $\bar{\epsilon}$ (0.03 to 0.06) (Figure 4d). However, for Cu$^{2+}$ concentration, the increases in $\bar{\alpha}$ (0.43 to 0.94) and $\bar{\epsilon}$ (0.02 to 0.24) are considerably larger (Figure 4e). Significant increases in $\bar{\alpha}$ (0.86 to 0.93) and $\bar{\epsilon}$ (0.02 to 0.17) are also observed with increasing reaction temperatures (Figure 4f).

The observations can be attributed to the scaled distributions of $d$ presented in Figure 4g-i, and to the increase in $h$ with the reaction parameters. As shown, the distribution does not change greatly with reaction time (Figure 4g), with mean diameter $\mu_d$ and standard deviation $\sigma_d$ showing only small rises. The corresponding spectral reflectances only show a slight absorption-broadening (Figure 4a). However, with concentration, the distribution of $d$ broadens drastically (Figure 4h), as does the absorption into the infrared (Figure 4b). Similar changes are observed for increasing reaction temperature (Figure 4i). The broadened distributions indicate an abundance of large nanoparticles ($d$ $\gtrsim$ 150 nm) in the layers. Since simulations show that the presence of such large particles leads to absorption-broadening into the infrared (Figure S3 of the SI), very large particles could account for the large increases in emittance in the infrared wavelengths for concentration (Figure 4e) and temperature (Figure 4f).

Layer thickness (*h*) may also play a role. As evidenced by the presence of large particles, higher concentrations and temperatures lead to thicker Cu depositions. Simulations show that an increasing *h* leads to an increasing $\bar{\epsilon}$ – particularly for small particles (*d* < 100 nm, as in Figures 3d, and S2 of the SI). And since the size distributions shown are all right-skewed, thicker layers have an abundance of such small particles in addition to large ones, and can show higher $\bar{\epsilon}$. From SEM images (SI, Section 5), the nanoparticle layer appears to have a few stacks of small and large particles. The thicknesses (*h*) are estimated to be between 200-400 nm – sufficient to show thickness-dependent optical performance, as reflected in Figures 3d-e (e.g. for $d \lesssim$ 100 nm and $h \gtrsim$ 150 nm).

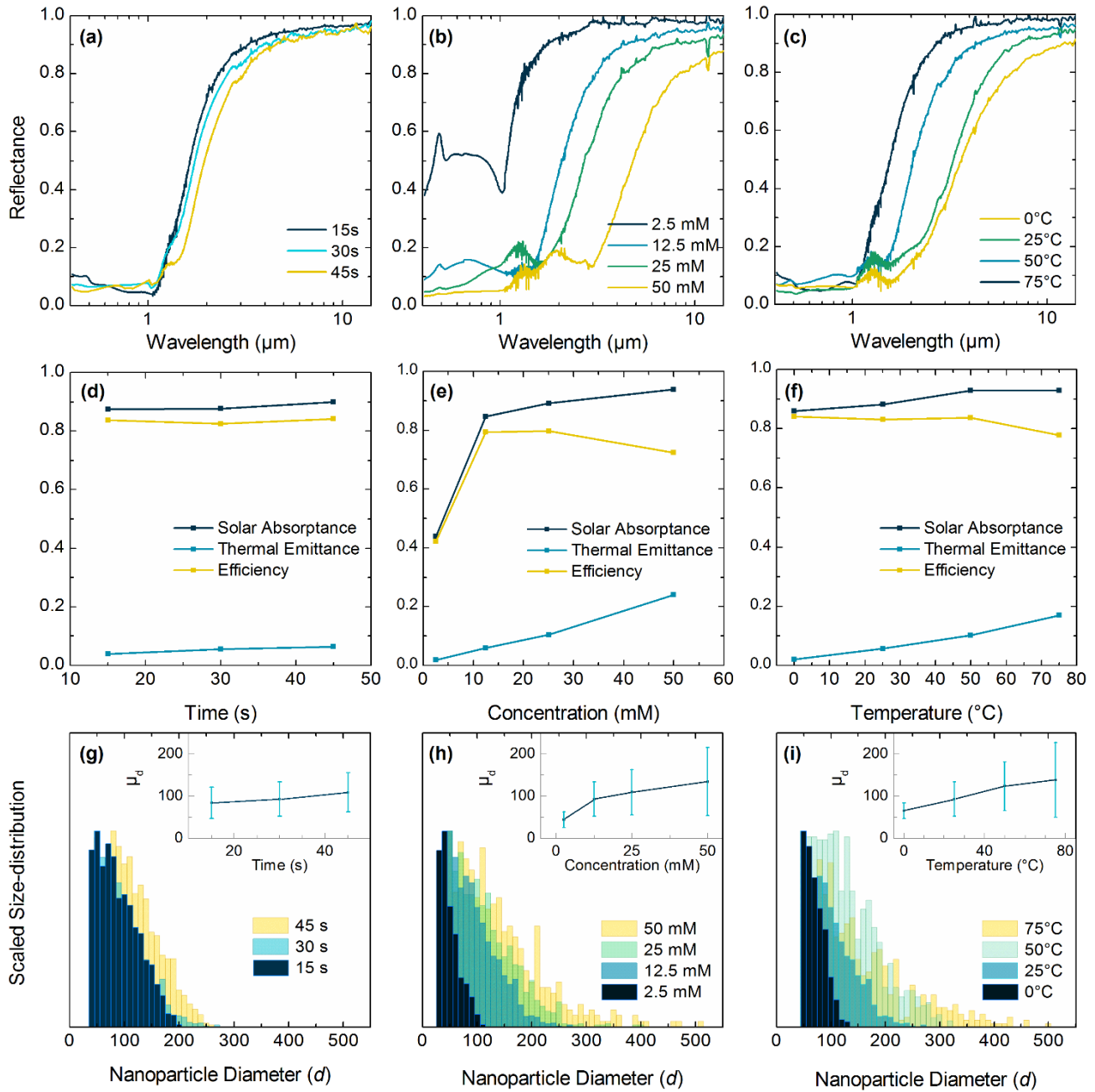

**Figure 4.** a-c) Spectral reflectances, d-f) $\bar{\alpha}$, $\bar{\epsilon}$ and $\eta_{sol-th}$, and g-i) scaled (to the same height) nanoparticle size-distribution of the PNFs fabricated with different times, temperatures and concentrations. Mean diameter $\mu_d$ and standard deviation $\sigma_d$ (error bars) of the distributions are presented in the insets.

It is evident from Figure 4d-f that the 'dip-and-dry' technique can be used to tune the PNFs' $\bar{\alpha}$, $\bar{\epsilon}$, and therefore, $\eta_{sol-th}$. For increasing temperatures and concentrations, the large accompanying increases in $\bar{\epsilon}$ outweigh the smaller increases in $\bar{\alpha}$, causing $\eta_{sol-th}$ to drop, whereas for immersion time, the small increases in $\bar{\alpha}$ and $\bar{\epsilon}$ maintain a near-constant value of $\eta_{sol-th}$. A peak $\eta_{sol-th}$ of 0.84, which is within the range predicted by the simulations, is observed for a $Cu^{2+}$ concentration of 12.5 mM, an immersion time of 45s, and a temperature of ~25°C. The parameters are used for subsequent experiments. The authors note that $\bar{\alpha}$ and $\bar{\epsilon}$ could be similarly tuned to maximize the PNFs' efficiency for different solar concentrations,[10] operating temperatures and radiative environments, or even attain the desired spectral reflectances for applications such as thermal heat detection (e.g. selective infrared reflectance[28]) – evidently, the 'dip-and-dry' technique is a broadly applicable technique for achieving such controllable selectivities.

Besides nanoparticle size distribution and layer thickness, surface texture can play an important role in determining the spectral reflectance of the PNF. The authors discovered that a stepwise 'dip-and-dry' process can alter the surface morphology of the nanoparticle layer. **Figure 5a** shows typical results for continuous 45s and stepped 30+15s depositions. As evident, stepwise depositions yield a lower $R(\lambda)$ at the measured wavelengths, and thus higher $\bar{\alpha}$ and $\bar{\epsilon}$ than those for the continuous case. The insets show that while the continuous depositions yield nanoparticles that resemble jagged spheroids, stepwise depositions yield additional, sharp, filament-like features that increase surface roughness. It is likely that these features reduce reflectance by providing a better optical impedance match between air and the nanoparticle layer, and by in-plane scattering of the incident radiation.[21, 29] For the PNF, this results in an average $\bar{\alpha}$ of 0.96 and $\bar{\epsilon}$ of 0.08. The values are comparable to notable results in the literature, and could potentially be enhanced further using antireflection coatings.[17, 22] The authors also note that the morphology of the PNF can be varied with the anions and surfactants in the solution,[30] type of metal deposited, and co- or serial deposition of different metals (SI, Section 6). The 'dip-and dry' technique therefore offers additional ways to tune the optical selectivity to those mentioned earlier.

Excellent wide-angle selective absorption is critical for SSAs in real situations, as the relative angle of the sun varies drastically during the day and across seasons (SI, Section 4). Furthermore, diffuse sunlight from the atmosphere constitutes a significant fraction, ~10%, of the incident solar power.[31] Therefore, to efficiently harvest solar energy, an SSA must have a high $\bar{\alpha}(\theta)$ and a low $\bar{\epsilon}$ at all incidence angles. The angle-dependence of $\bar{\alpha}$ and $\bar{\epsilon}$ is not well-studied, and reports of designs with $\bar{\alpha}(80°) > 0.7$ are unfortunately rare due to reflection at high angles. Here, the authors found that PNFs yielded by the stepwise 'dip-and-dry' technique have an excellent wide-angle solar absorptance, with $\bar{\alpha}(\theta)$ ranging from 0.96 at 15°, to a peak of 0.97 at ~35°, to 0.79 at 80°. This insensitivity to angle is likely due to the spheroidal particles and filament-like features, which should provide near-constant optical responses over a wide range of incidence angles. In contrast, wide-angle SSAs reported in literature often have directionally orientated structures, which likely enhances near-normal $\bar{\alpha}(\theta)$, but reduces $\bar{\alpha}(\theta)$ at high angles.[17, 19, 29, 32] As shown in Figure 5b, for $\theta > 50°$, the PNFs have a significantly higher $\bar{\alpha}(\theta)$ than SSAs with comparable optical selectivities, which makes them attractive as both tubular and flat SSAs. At very high values of $\theta$, however, $\bar{\alpha}(\theta)$ falls – likely due to Mie-scattering back into free-space. For $\bar{\epsilon}$, an increase with angle is observed (0.08 at 15° to 0.12 at 80°). This is expected, since at higher angles of incidence, the optical paths through the attenuating nanoparticle are longer. The increased emittance may also arise from the underlying Zn, which, being metallic, has higher emittance at large angles.[33] The angular measurements yield an upper bound of 0.10 for the hemispherical $\bar{\epsilon}$. The efficiency $\eta_{sol-th}$ is found to peak at ~0.86 – a desirable value for selective solar absorption.

Theoretically, the high $\eta_{sol-th}$ of the PNFs allows them to attain higher temperatures under the sun than those of commonly found black, emissive materials. Laboratory simulations of such a scenario confirms this to be the case: under illumination by a solar simulator lamp with irradiance equivalent to '1 Sun', the PNFs are found to reach a significantly higher temperature (148°C), than that (122°C) of a thermally emissive black aluminum foil ($\bar{\alpha}$ = 0.95 and $\bar{\epsilon}$ ~ 0.75). Notably, the difference in temperature is achieved with the radiative environment of the samples representing a very warm 'sky' (effective blackbody temperature of the lab environment > 30°C) (SI, Section 7).

Under real skies with effective radiative temperatures ≲ -30°C, the difference would likely be ≳40°C, owing to significantly increased radiative heat losses from the black foil.

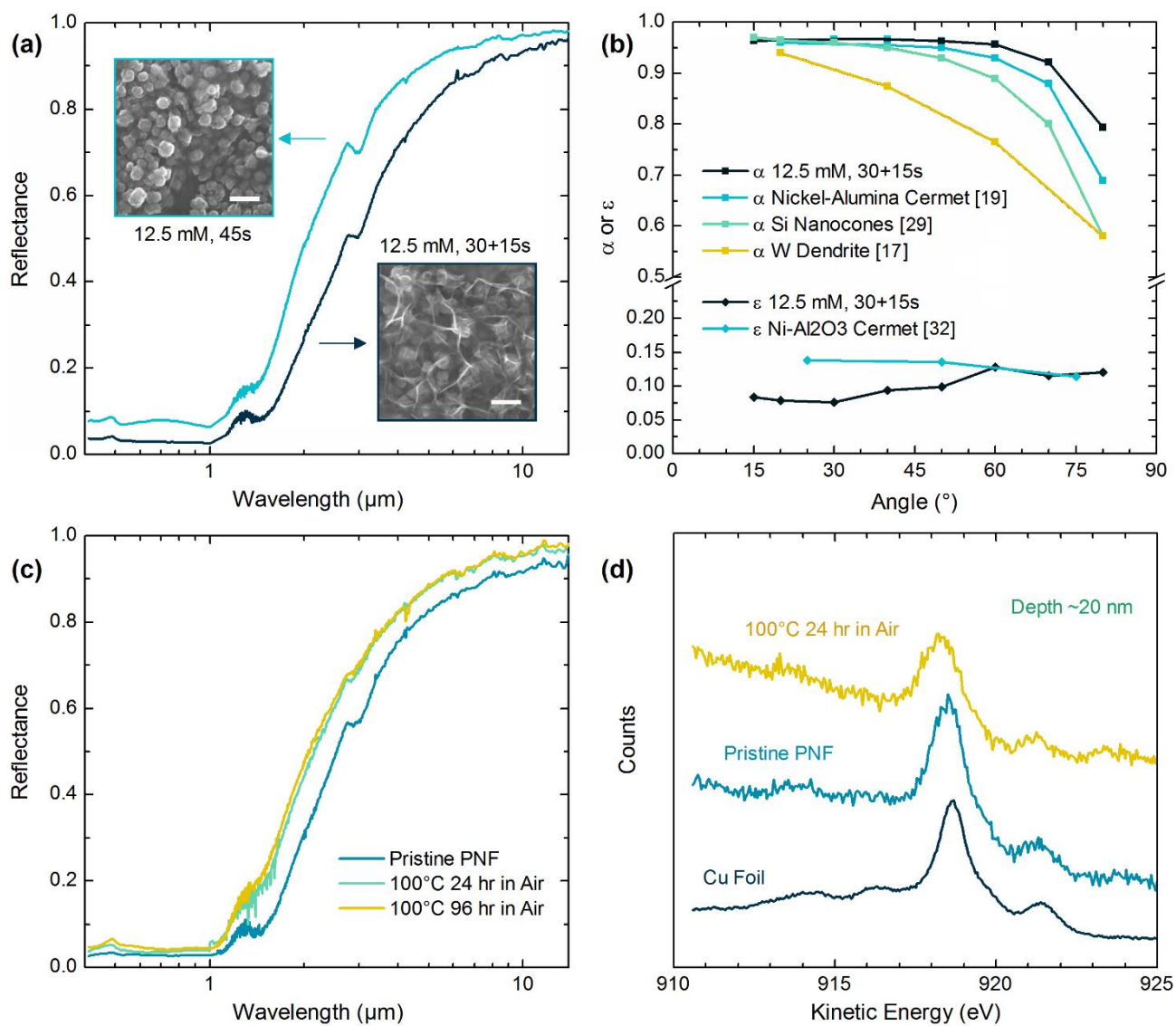

**Figure 5.** a) Reflectance spectra of PNFs fabricated using continuous (45s) and stepwise (30+15s) deposition. Insets show the SEM images of the deposited nanoparticles, with scale bars representing a length of 200 nm. b) Angular reflectance and emittance of the PNFs fabricated by stepped deposition, presented alongside those of different solar absorptive surfaces reported in the literature. c) Reflectance spectra of PNFs before, after 24 hours and after 96 hours of heating in air at 100 °C. d) Auger spectra of pristine and heat-treated PNFs taken at ~20 nm depth below the surface, and of a reference Cu foil that had surface oxide removed.

For practical use, an SSA needs to retain its performance under operating conditions over a prolonged period. Therefore, it must be mechanically and thermally stable. In that regard, PNFs subjected to mechanical and thermal stability tests show promising results. The mechanical stability tests are conducted following the ASTM adhesion test (D3359-09) protocol: a standard tape with known adhesiveness is procedurally applied to PNFs, and then peeled off. It is observed that the adhesion between the nanoparticle layer and the Zn substrate is strong enough for the adhesives of the tapes to remain on the PNFs (SI, Figure S10). The authors conjecture that this could be due to the alloying of Cu and Zn at the nanoparticle-substrate interface. In one case, where the tape was completely peeled off from a PNF, reflectance measurements yielded no significant changes in $\bar{\alpha}$, $\bar{\epsilon}$ and $\eta_{sol-th}$ (**Table 1**). The test results are encouraging, and indicate the PNFs' compatibility with post-deposition rolling, cutting, and other manufacturing processes.

To test for thermal stability, bare PNFs are subjected to accelerated thermal aging tests at 200°C in argon and air for up to 96 hours. Analogous tests are also performed in targeted operating conditions (100°C, in air). Figure 5c and Table 1 show typical results. In both argon and air, $\bar{\alpha}$ and $\bar{\epsilon}$ decrease slightly with prolonged heating. For instance, over 96 hours of heating in air at 200°C, $\bar{\alpha}/\bar{\epsilon}$ decrease from 0.94/0.13 to 0.90/0.09. Regardless, $\eta_{sol-th}$ remains essentially unchanged in all cases, and in air, the performance stabilizes within 24 hours. Furthermore, the results for argon and air respectively indicate that any Cu-Zn alloying and oxidation do not significantly impact the PNFs' performance. In fact, even after prolonged heating at the targeted 100°C operating temperature, the PNF's surface morphology remains intact, and AES measurements indicate that any oxidation is limited to a depth ≲ 20 nm from the surface – small compared to the thickness of the nanoparticle layer (~200-400 nm) (Figures 5d, and S11b of the SI). XRD measurements of Cu particles heated at 100°C also show pure copper peaks (Figure S11a). The observed thermal stability could be due to the presence of reducing Zn, which underlies and also forms a thin layer around the Cu nanoparticles (Figure 2e). Furthermore, the stabilities observed at 200°C suggest that the PNFs may be operable at temperatures much higher than the 100°C value conservatively surmised by the authors. For harsh environments, the stability may be further enhanced by applying impermeable anti-reflection coatings.[34] For vacuum-tube SSA configurations, however, such additions would be redundant.

The 'dip-and-dry' technique discussed above is a simple approach to fabricate plasmonic SSAs with high performance. It is therefore surprising that no precedent to our use of this technique was found in literature. An apparently similar approach was reported by Banerjee et. al.,[35] who used the Cu-Zn galvanic displacement reaction as a prelude to a high-temperature oxidation step for obtaining a previously known semiconductor-metal tandem SSA (CuO microparticles on metal (Fe)).[36] The CuO-Fe tandem, however, has a fundamentally different solar absorption mechanism (interband electron transitions) compared to plasmonic processes, and its performance ($\bar{\alpha} \leq 0.91$ and $\bar{\epsilon} \geq 0.17$), has long been superseded by more recent designs.[10, 18-19, 22, 24] In contrast, the PNF presented here is a plasmonic SSA fabricated by a single-step, galvanic-displacement based, room-temperature 'dip-and-dry' technique. And as demonstrated, its optical performance ($\bar{\alpha}$ ~ 0.96 and hemispherical $\bar{\epsilon}$ < 0.10) is superior to Banerjee et. al.'s design, and on par with current SSA architectures.[10, 18-19, 23-24]

**Table 1**. Solar absorptance, thermal emittance and solar-thermal conversion efficiencies of PNFs before and after mechanical and thermal stability tests.

| Time (hr) | 200°C in Argon | | | 200°C in air | | | 100°C in air | | | Mechanical Stability Test | | | |
|---|---|---|---|---|---|---|---|---|---|---|---|---|---|
| | $\bar{\alpha}$ | $\bar{\epsilon}$ | $\eta_{sol-th}$ | $\bar{\alpha}$ | $\bar{\epsilon}$ | $\eta_{sol-th}$ | $\bar{\alpha}$ | $\bar{\epsilon}$ | $\eta_{sol-th}$ | | $\bar{\alpha}$ | $\bar{\epsilon}$ | $\eta_{sol-th}$ |
| 0 | 0.94 | 0.10 | 0.83 | 0.94 | 0.13 | 0.79 | 0.95 | 0.09 | 0.85 | Before | 0.93 | 0.09 | 0.83 |
| 24 | 0.93 | 0.07 | 0.85 | 0.90 | 0.09 | 0.81 | 0.92 | 0.05 | 0.87 | After | 0.94 | 0.08 | 0.85 |
| 96 | 0.91 | 0.08 | 0.83 | 0.90 | 0.09 | 0.80 | 0.91 | 0.05 | 0.86 | | | | |

In summary, a room-temperature, 'dip-and-dry' technique for fabricating plasmonic nanoparticle-coated foils (PNFs) for selective solar absorption is demonstrated. The PNFs yielded by the technique not only exhibit an excellent, wide-angle solar absorptance $\bar{\alpha}$ (0.96 at 15°, to 0.97 at 35°, to 0.79 at 80°) and a low hemispherical thermal emittance $\bar{\epsilon}$ (< 0.10) without the aid of antireflection coatings, but also show mechanical strength and thermal stability at temperatures of up to 200°C in air over prolonged periods. Furthermore, and importantly, the fabrication process is simple, inexpensive and environment-friendly compared to commercial SSA manufacturing methods, and allows for convenient tuning of the PNFs' solar absorptance and thermal emittance. The authors therefore propose this as an appealing alternative to current SSA fabrication techniques.

**Experimental Section**

*Fabrication of the Plasmonic Nanoparticle-coated Foils (PNFs):* Zn foil (thickness – 250 µm, purity – 99.98%, from Alfa Aesar) was cut and flattened into small strips. Prior to Cu deposition, the strips were sonicated in 2.5% sulfuric acid, IPA, and acetone for 30 seconds, 5 minutes, and 5 minutes respectively. After rinsing with IPA and water and blow-drying with air, the Zn strips were immersed into aqueous $CuSO_4$ with different concentrations, at different temperatures, or for different times for the nanoparticles to form. The samples were then dipped in water to quench the reaction and then blow dried. Similar experiments were also performed with Zn-coated Al foils, and yielded optically selective PNFs (SI, Section 9).

*Optical characterization:* Spectral reflectance of the PNFs was determined separately in the visible to near-infrared (0.41-1.05 µm) and near-infrared to mid-infrared (1.06-14 µm) wavelength ranges. For the first range, a high-power supercontinuum laser (SuperK Extreme, NKT Photonics) coupled to a tuneable filter (Fianium LLTF contrast) was used to shine specific wavelengths into an integrating sphere (Model IS200, Thorlabs). PNFs were then individually inserted into the integrating sphere to intercept the light at angles of incidence between 15° and 80°, and had their reflectances measured at 5 nm wavelength intervals by a silicon detector. For the near-infrared to mid-infrared range, a Fourier Transform Infrared (FT-IR) spectrometer (Vertex 70v, Bruker) and a gold integrating sphere (Model 4P-GPS-020-SL, Labsphere), along with a mercury-cadmium telluride detector were similarly used. In both cases, gold-coated aluminum foils and gold-coated silicon wafers were used as references suited to the PNFs, which themselves showed nearly-specular visible reflectances. For PNFs fabricated under the same conditions, averaged spectral reflectances of multiple samples were patched and extrapolated to the 280 nm – 40 μm range as described in the SI (Section 2). $\bar{\alpha}$, $\bar{\epsilon}$ and $\eta_{sol-th}$ were calculated from the resultant spectrum using Equations (1)-(3).

*Material Characterizations:* Auger Electron Spectroscopy (AES) was performed on one pristine and one thermally annealed (100°C, 24 hours) PNF made using the stepwise deposition process. The spectra were obtained for different depths from the PNF's surfaces by etching with an argon ion beam. A copper foil, with surface oxide etched, was used as reference. Transmission electron microscopy, electron diffraction, EDS mapping and X-ray diffraction measurements were taken on Cu nano- and micro-particles. The particles were obtained by sonicating PNFs with thick copper depositions, followed by cleaning the particles with water, and then dispersal and prolonged sonication in acetone to form a dispersion.

*Simulation:* FDTD Solutions 8.6.1 software by Lumerical was used to simulate the spectral reflectance of PNFs with various nanoparticle diameter *d* and layer thicknesses *h*. The Cu nanoparticles, which appear as comprising of smaller structures, were represented as spherical clusters of 20 nm spheres in a face centered cubic arrangement. The clusters were randomly placed in layers over Zn, which was represented as a smooth, optically thick slab. The volume fraction of the nanoparticles was chosen to be ~0.55 based on analysis of SEM images of preliminary samples and the assumption that any loose packing due to random nanoparticle growth would be compensated by the close-packing of nanoparticles with different sizes. Depending on *d,* it varied between 0.52 and 0.57. Refractive index data for Cu on Zn were obtained from the literature.[37-39] From the simulations, spectral reflectance was obtained and used to calculate and $\bar{\alpha}$, $\bar{\epsilon}$ and $\eta_{sol-th}$ at normal incidence. The stated values of *h* were accurate to ± 5%, as the particles did not stack up to precise heights.

*SEM characterization:* Cu nanoparticles on the PNFs were imaged with a Zeiss Sigma VP scanning electron microscope. For PNFs synthesized under different parameters, the sizes of nanoparticles in the layers imaged from the top were measured using ImageJ. From the measurements, the particle size distributions, means and standard deviations were calculated for the different reaction conditions.

*Solar Simulator Test:* A PNF, placed atop a thermally insulating foam, was put in a white, open-top box with a polyethene cover to reduce convection and allow transmission of solar and thermal radiation. The setup was then placed directly underneath a solar simulator lamp (Atlas SolarConstant, with K.H. Steuernagel Lichttechnik GmbH controller) with intensity adjusted to that of the AM 1.5 Global Solar Spectrum. Insulating foam was placed around the setup, which was then allowed to reach steady state. Temperatures of the sample and the environment were recorded using thermocouples. For comparison, an analogous test was performed with aluminum foil coated with a black, thermally emissive layer (item 7073T24, McMaster Carr). Additional details are available in the SI Information (section 7).

*Mechanical stability tests:* The ASTM adhesion test (D3359-09) protocol was followed, but without making cuts on the PNFs – Elcometer 99 tape was pressed and smoothed onto the PNFs whose spectral reflectances had been previously measured, and after 90 seconds, removed rapidly at right angles to the samples.

*Thermal Stability Tests:* Individual PNFs were heated on hotplates for 96 hours, either in an argon-filled chamber or in air, at 100°C or 200°C. During the tests, the PNFs were covered with a lid to maintain a uniform temperature. Spectral reflectances of the samples were measured prior to the test, and after heating for 24 and 96 hours, to measure the change in $\overline{\alpha}$, $\overline{\epsilon}$ and $\eta_{sol-th}$ over time.

**Supporting Information (SI)**
Supporting Information is available from the Wiley Online Library or from the authors.

**Acknowledgements**
The work was supported by startup funding from Columbia University, NSF IGERT program (grant # DGE-1069240), AFOSR MURI (Multidisciplinary University Research Initiative) program (grant # FA9550-14-1-0389), and AFOSR DURIP (Defense University Research Instrumentation Program) program (grant # FA9550-16-1-0322). The authors would also like to thank Cheng-Chia Tsai of the Department of Applied Physics at Columbia University for his help on this study and Sagar Mandal of the Department of Computer Engineering at Georgia Institute of Technology for guidance on figure design.